\documentclass[aps,prc,reprint,showpacs]{revtex4-1}

\usepackage{amsmath}
\usepackage{graphicx}
\usepackage{bm}
\usepackage[dvips]{color}

\usepackage{txfonts}
\usepackage[colorlinks=true,linkcolor=blue,citecolor=blue,urlcolor=blue]{hyperref}

\hypersetup{
  pdftitle={Universality of short-range nucleon-nucleon correlations},%
  pdfauthor={Thomas Neff <t.neff@gsi.de>}}

\newcommand{\op}[1]{\hat{#1}}
\renewcommand{\vec}[1]{\bm{\mathit{#1}}}

\newcommand{\bra}[1]{\ensuremath{\langle{#1}|\,}}
\newcommand{\ket}[1]{\ensuremath{\,|{#1}\rangle}}
\newcommand{\braket}[2]{\ensuremath{\langle{#1}|{#2}\rangle}}
\newcommand{\matrixe}[3]{\ensuremath{\langle{#1}|\,{#2}\,|{#3}\rangle}}
\newcommand{\expect}[2]{\matrixe{#1}{#2}{#1}}

\newcommand{\fm}{\ensuremath{\textrm{fm}}}
\newcommand{\nuc}[2]{$^{#1}${#2}}

\newcommand{\CO}{\ensuremath{\op{C}}}
\newcommand{\COd}{\ensuremath{\op{C}^{\dagger}}}
\newcommand{\Rm}{\ensuremath{R_-}}
\newcommand{\DRm}{\ensuremath{R'_-}}

\begin{document}

\title{Universality of short-range nucleon-nucleon correlations}

\author{H. Feldmeier}
\author{W. Horiuchi}
\altaffiliation[Present adress: ]{RIKEN Nishina Center, Wako 351-0198, Japan}
\author{T. Neff}
\affiliation{GSI Helmholtzzentrum f\"ur Schwerionenforschung GmbH,
Planckstra\ss e 1, 64291 Darmstadt, Germany}

\author{Y. Suzuki}
\affiliation{Department of Physics, Niigata University, Niigata 950-2181, Japan}
\affiliation{RIKEN Nishina Center, Wako 351-0198, Japan}

\begin{abstract} 
  Short-range correlations between nucleon pairs in different spin-isospin channels are investigated for light nuclei using the Argonne~v8$^\prime$ interaction. At distances below 1~fm a universal behavior is found for the deuteron, \nuc{3}{H}, \nuc{3}{He} and for ground and first excited states in \nuc{4}{He}. This behavior in coordinate space is reflected by a universal behavior for the high-momentum components in momentum space. The universality indicates that a pairwise renormalization is possible in order to obtain a universal effective two-body interaction that does not scatter to high momentum states. The exact two-body densities are compared with those obtained using the unitary correlation operator method with simple trial wave functions. The effect of three-body correlations due to the tensor force on the two-body densities is discussed.
\end{abstract}


\maketitle

\section{Introduction} 
\label{sec:intro}
 
Realistic nucleon-nucleon interactions, which reproduce the scattering phase shifts, imply usually strong repulsive and tensor forces at short distances. These induce short-range correlations in the nuclear many-body system which complicate the theoretical description so much that exact solutions of the many-body Schr{\"o}dinger equation become unfeasible for systems with more than about twelve nucleons. Therefore theoretical methods have to be devised in order to tackle this problem.

At short distances, where the scattering nucleons overlap strongly, there is no unique way to parametrize the complex many-body quantum chromodynamics problem in terms of just the distance, the relative momentum, and the spins of the nucleons. In all models for the nucleon-nucleon interaction the short-range behavior is governed by form factors of various types without rigorous derivations. Therefore experimental data for elastic scattering, which provide phase shifts for energies up to the pion threshold cannot sufficiently constrain the nucleon-nucleon potential at small distances. In consequence, different phase shift equivalent interactions, (e.g., \cite{wiringa95,machleidt01,entem03,epelbaum05}), show a quite different high-momentum or short-range behavior.  

Another uncertainty arises when going from the two-nucleon scattering states to bound many-body states of nuclei. In the scattering situation the two nucleons are in an energy eigenstate with a well-defined relation between momentum or kinetic energy, potential energy, and total energy, usually labeled with on-shell. In the many-body case a nucleon pair, which interacts with the surrounding other nucleons, neither has sharp energy nor is their relative momentum related to their energy. In this situation the so-called off-shell behavior (i.e., local versus momentum-dependent parts) of the nuclear interaction is important but also not fully constrained by scattering data. This ambiguity in the off-shell behavior of two-body interactions is also related to the three-body interactions that should accompany the different two-body interactions \cite{polyzou90}.

Information about the short-range behavior of the nuclear interaction is contained in the one- and two-body momentum distributions of nucleons in finite nuclei. However, it is difficult to relate measured cross sections with momentum distributions. Reactions and their kinematic conditions have to be chosen properly, for example, in such a way that the correlated nucleon is removed instantaneously and final-state interactions are minimized \cite{frankfurt08,arrington11}. In recent years we have seen a renewed interest in studying short-range correlations using $(e,e'pp)$ and $(e, e'pn)$ \cite{egiyan06,baghdasaryan10} $(p,pp)$ and $(p, ppn)$ \cite{piasetzky06} experiments. One important result from these studies is that the momentum distributions above the Fermi momentum are dominated by tensor correlations \cite{subedi08}. There are also attempts to explore the effect of tensor correlations in nuclei by pick-up in $(p,d)$ reactions \cite{tanihata10}.

Over the years short-range correlations have been studied using approaches such as the coupled-cluster method \cite{zabolitzky78}, correlated basis functions \cite{co94,fabrocini00}, Green's functions \cite{dickhoff04}, variational methods \cite{pieper92} or within a cluster expansion approach \cite{alvioli05,alvioli08}. For a review see Ref.~\cite{muether00}. These methods were essentially applied only to doubly-close shell nuclei like \nuc{16}{O} and \nuc{40}{Ca}. For lighter systems pioneering studies have been performed in the Green's function Monte Carlo approach \cite{forest96,schiavilla07,wiringa08} employing two- and three-body interactions.

In this paper we do not investigate different realistic interactions but concentrate on the Argonne v8$^\prime$ (AV8$^\prime$) potential \cite{pudliner97} where the short-range physics is described by a phenomenological local potential. Extending on the results obtained in Ref.~\cite{suzuki09} the aim of this investigation is to solve the three- and four-nucleon system exactly and analyze the short-range correlations in the different spin-isospin channels. After explaining briefly the many-body approach in Sec.~\ref{sec:correlated} we define explicitly various one- and two-body densities that are used in Sec.~\ref{sec:results} to illustrate that the short-range pair correlations are universal in the sense that they depend very little on the surrounding  nucleons and the many-body state in general. This feature has been realized some time ago by Forest \textit{et al.} \cite{forest96}. Here we investigate the universality and discuss the implications to devise effective low-momentum interactions.

The universal behavior gives hope that it is possible to derive effective
interactions that are phase-shift equivalent and soft enough to permit many-body calculations with a Slater determinant basis without inducing large many-body effective potentials. The transformation of the Hamiltonian to an effective one implies of course also the same transformation of any other observable. In order to keep the physics transparent it is highly desirable that these operators, which are usually one-body operators, do not take on large many-body contributions when they are transformed to effective ones. This will be the case if the induced correlations are of short range and the theoretical treatment takes account of this, as in the unitary correlation operator method (UCOM). Many observables like radii or multipole moments are not sensitive to short-range correlations, however, observables containing the spins, as in Gamow-Teller transitions are affected by the tensor correlations.

In Sec.~\ref{sec:model} we briefly recapitulate the many-body method adopted here, define the one- and two-body densities in coordinate and momentum space, and discuss the AV8$^\prime$ potential in the different spin-isospin channels. In Sec.~\ref{sec:results} we display the correlations in coordinate and momentum space for the different spin-isospin-channels and investigate quantitatively their universality. In Sec.~\ref{sec:threebody} we discuss how three-body correlations manifest themselves in the two-body densities and in Sec.~\ref{sec:ucom} we compare to two-body densities obtained with the UCOM approach. Summary and outlook are drawn in Sec.~\ref{sec:summary}.

\section{Many-body model, interactions, densities}
\label{sec:model}

\subsection{Correlated Gaussian basis approach}
\label{sec:correlated}

We assume that an $A$-nucleon state can be expanded in terms of a
combination of basis states, each of which is a product of space, spin
and isospin parts,
\begin{equation}
  \ket{\Psi;JM} =
	\sum_{i=1}^K C_i{\cal A}\Big\{
    \left[\ket{\psi_i^{\rm (space)} \psi_i^{\rm (spin)}}\right]_{JM}
    \ket{\psi_{i}^{\rm (isospin)}} \Big\} \: .
  \label{basis.LS}
\end{equation}
Here ${\cal A}$ is the antisymmetrizer and the square bracket
$[\cdots]$ stands for the angular momentum coupling. The spin and
isospin parts are expanded using the basis of successive coupling,
e.g.,
\begin{equation}
  \ket{\psi_{i}^{(\rm spin)}}=\big|[\cdots [[[\textstyle{\frac{1}{2}}
          \textstyle{\frac{1}{2}}]_{S_{12}}
        \textstyle{\frac{1}{2}}]_{S_{123}}]\cdots]_{S_iM_S}\rangle,
  \label{basis.spin}
\end{equation}
where the set of intermediate spins $(S_{12},S_{123},\ldots)$ takes
all possible values compatible with the total spin $S_i$ of the $i$th
basis. The isospin mixing is ignored in this paper, so that the total
isospin $T_i$ is kept fixed to $T$.  The orbital part $\psi_{i}^{(\mathrm{space})}$ is expressed in terms of the explicitly correlated Gaussian
basis~\cite{varga95,suzuki08} as explained below.

We denote the position operator of particle $i$ by $\op{\vec{r}}_i$. To simplify the notation, we define this position to be
measured from the total center of mass of the system. The correlated
Gaussian basis is conveniently expressed in terms of the relative
coordinates, e.g., the Jacobi set of the coordinates, $\op{\vec{x}}=(\op{\vec{x}}_1,\ \op{\vec{x}}_2,\ldots, \op{\vec{x}}_{A-1})$: $\op{\vec{x}}_1=\op{\vec{r}}_1-\op{\vec{r}}_2,\ \op{\vec{x}}_2=(\op{\vec{r}}_1+\op{\vec{r}}_2)/2 -\op{\vec{r}}_3,\ldots$ The correlated Gaussian evaluated at
the position $\vec{x}$ corresponding to the operator $\op{\vec{x}}$
takes the following form
\begin{equation}
  \braket{\vec{x}}{\psi_i^{\mathrm{(space)}}}= {\rm
    exp}\Big(-{\frac{1}{2}}{\widetilde{{\vec{x}}}} A_i {\vec{x}}\Big)
  \left[{\cal Y}_{L_{1i}}({\widetilde{u_i}}{\vec{x}}) {\cal
      Y}_{L_{2i}}({\widetilde{v_i}}{\vec{x}})\right]_{L_iM_L}\ ,
  \label{cg2}
\end{equation}
where ${\cal Y}_{LM}({\widetilde{u}}{\vec{x}})=|{\widetilde{u}}{\vec{x}}|^{L} Y_{LM}({{\widetilde{u}}{\vec{x}}})$ is a solid spherical
harmonics.  In Eq.~(\ref{cg2}) the matrix notation is used to simplify
the expression. That is, $u_i$ is an $A-1$ dimensional column vector
and $\widetilde{u_i}$ denotes its transpose, $\widetilde{u_i}{\vec{x}}=\sum_{j=1}^{A-1}(u_{i})_j{\vec{x}}_j$.  Similarly, $A_i$ is an
$(A-1)\times (A-1)$ positive-definite, symmetric matrix, and
${\widetilde{{\vec{x}}}} A_i {\vec{x}}$ is a short-hand notation for
$\sum_{j,k=1}^{A-1}(A_{i})_{jk}{\vec{x}}_j\cdot {\vec{x}}_k$. The basis is
in fact correlated because all the coordinates are coupled through the
off-diagonal elements of $A_i$.  The elements of $A_i$ and $u_i$ (and
$v_i$) are parameters to characterize the ``shape'' of the basis
function.

The matrix elements of the Hamiltonian can be analytically obtained
using the generating function technique.  All the formulas needed are
given in Refs.~\cite{suzuki08,suzuki09,aoyama11}.  As seen in
Eqs.~(\ref{basis.LS}), (\ref{basis.spin}) and (\ref{cg2}), each basis
function contains both discrete and continuous parameters. The former
includes $L_{1i}, L_{2i}, L_i, S_{12}, S_{123}, \ldots, S_i, T_{12},
T_{123}, \ldots, (T_i=T)$ and the latter the elements of $A_i, u_i$,
and $v_i$.
Though the Gaussians may not be ideal to cope with the repulsion, it
is in fact possible to obtain results as accurate as other
sophisticated methods for a few-body system~\cite{kamada01,suzuki08}. 
One of the advantages of the present
method is that the state $\Psi$ is expressed analytically so that
physical quantities of interest can be readily evaluated. Since the
Fourier transform of the correlated Gaussian basis is also expressed
as correlated Gaussians in momentum variables~\cite{suzuki08,aoyama11}, it is
straightforward to calculate the matrix element of a quantity
depending on the momentum operator.  To have a compact basis size $K$
saves time of computations.  We use the stochastic variational
method~\cite{varga95,suzuki:svm} to choose the parameters and increase
the basis dimension until good convergence is reached.

\subsection{One- and two-body density}

The antisymmetrized many-body state $\ket{\Psi;JM}$ contains all the
information about the nuclear system.  For example, the one-body
density in coordinate space is defined as
\begin{equation} 
  \rho^{(1)}(\vec{r}_1)=\frac{1}{2J+1}\sum_M
  \expect{\Psi;JM}{\sum_{i=1}^A
  \delta^3(\op{\vec{r}}_i-\vec{r}_1) } \ ,
\end{equation}
where $\op{\vec{r}}_i$ is the position operator for the $i$th particle
measured from the position of the total center of mass.  Likewise the
one-body momentum distribution is calculated as
\begin{equation} 
  n^{(1)}(\vec{k}_1)=\frac{1}{2J+1}\sum_M
  \expect{\Psi;JM}{\sum_{i=1}^A \delta^3(\op{\vec{k}}_{i}-\vec{k}_1) } \ ,
\end{equation}
where the momentum $\op{\vec{k}}_i$ of particle $i$ is defined in the
total momentum frame of the nucleus. That means if a particle has
momentum $\vec{k}_1$ all other particles have together a total momentum
$-\vec{k}_1$.

One should keep in mind that the possibility of finding a single
nucleon with momentum $\vec{k}_1$ does not imply that this nucleon has
an energy that is related to $\vec{k}_1$, such as $\vec{k}_1^2/(2m_N)$.
As all nucleons are interacting with each other one can not define an
observable for the energy of one nucleon.

In a mean field picture, where particles move independently in a
common single-particle potential, each particle can be assigned to a
single-particle state that has a sharp energy, the single-particle
energy.  But this energy is also not uniquely related to a momentum,
because the state has spread-out distributions in momentum and in
coordinate space which are related.

Similar effects occur in the interacting many-body case where only the
total energy (i.e., the eigenvalue of the Hamiltonian) is well
defined. Rapid spatial variations in the many-body wave function show
up as increased probabilities at large single-particle momenta. For
example strongly repulsive two-body interactions induce areas where
one will not find particle pairs because their interaction energy
would be large and positive.  At the edges of these correlation holes
the wave function has to vary rapidly giving rise to large momenta and
extra positive kinetic energy. But altogether it is more profitable to
pay the kinetic energy and avoid the even larger potential energy by
staying out of the repulsive region. Thus the high-momentum tail of
the momentum distribution reflects the short-range correlations.

In Sec.~\ref{sec:results} we discuss these phenomena by looking at different exact many-body eigenstates. The one-body densities can be accessed in scattering experiments, the proton density preferably by electron scattering.

A more direct way to see short-range correlations is given by the
two-body density  
\begin{multline}  \label{eq:rhorelr12}
  \rho^{(2)}_{SM_S,TM_T}(\vec{r}_1,\vec{r}_2)= \frac{1}{2J+1} \times \\
  \sum_M \expect{\Psi;JM}{\sum_{i<j}^A
    \op{P}^{SM_S}_{ij}\op{P}^{TM_T}_{ij} \delta^3(\vec{\op{r}}_i-\vec{r}_1)\
    \delta^3(\vec{\op{r}}_j-\vec{r}_2)} \ ,
\end{multline}
where $\rho^{(2)}_{SM_S,TM_T}(\vec{r}_1,\vec{r}_2)$ is the probability
density that a nucleon pair with one nucleon at position $\vec{r}_1$
and the other one at $\vec{r}_2$ is found in the spin $S,M_S$ and
isospin $T,M_T$ channel.  $\vec{r}_1$ and $\vec{r}_2$ are measured from
the total center-of-mass position and $\op{P}^{SM_S}_{ij}$ and
$\op{P}^{TM_T}_{ij}$ project on spin and isospin of the pair,
respectively. The label $T$ here indicates the two-nucleon
isospin. Note that it is also used to denote the total isospin of the
system in the previous subsection.

To keep the graphical presentation transparent we discuss the
short-range correlations as a function of the relative position
$\vec{r}\equiv \vec{r}_1-\vec{r}_2$ of the two nucleons only and
integrate over their center-of-mass position $\vec{R}\equiv (\vec{r}_1+\vec{r}_2)/2
$.
\begin{multline}
  \rho^{\rm{rel}}_{SM_S,TM_T}(\vec{r}) = \frac{1}{2J+1} \times \\
	\sum_M \expect{\Psi;JM}{\sum_{i<j}^A \op{P}^{SM_S}_{ij}
    \op{P}^{TM_T}_{ij} \delta^3(\vec{\op{r}}_i-\vec{\op{r}}_j-\vec{r})} \ .
  \label{eq:rhorelr}
\end{multline}
The corresponding distribution of the relative momentum
$\vec{k} \equiv (\vec{k}_1-\vec{k}_2)/2$ of the particle pair with total spin
$S,M_S$ and isospin $T,M_T$ is defined as
\begin{multline}
  n^{\rm{rel}}_{SM_S,TM_T}(\vec{k}) = \frac{1}{2J+1} \times \\
	\sum_M \expect{\Psi;JM}{\sum_{i<j}^A \op{P}^{SM_S}_{ij} \op{P}^{TM_T}_{ij}
    \delta^3\big(\frac{1}{2}(\vec{\op{k}}_i-\vec{\op{k}}_j)-\vec{k}\big)} \ .
  \label{eq:rhorelk}
\end{multline}
We also define the two-body densities $\rho^{\rm{rel}}_{S,T}(r)$ and $n^{\rm{rel}}_{S,T}(k)$ that are obtained by summing the spin- and isospin-indices $M_S$ and $M_T$ so that they do not depend on the orientation of $\vec{r}$ and $\vec{k}$
\begin{equation}
	\rho^{\rm{rel}}_{S,T}(r) = \sum_{M_S,M_T} \rho^{\rm{rel}}_{SM_S,TM_T}(\vec{r})
\end{equation}
and
\begin{equation}
	n^{\rm{rel}}_{S,T}(k) = \sum_{M_S,M_T} n^{\rm{rel}}_{SM_S,TM_T}(\vec{k}) \: .
\end{equation}

The distributions defined in Eqs.~(\ref{eq:rhorelr}) and (\ref{eq:rhorelk}), when coupled properly in space-spin space, are called internucleon correlation functions in Ref.~\cite{suzuki09}. The internucleon correlation functions contain all the information needed to calculate the energy of the state for a two-body Hamiltonian.

\subsection{Realistic nuclear forces}
\label{sec:forces}

\begin{figure*} 
  \includegraphics[angle=0,width=0.85\textwidth]{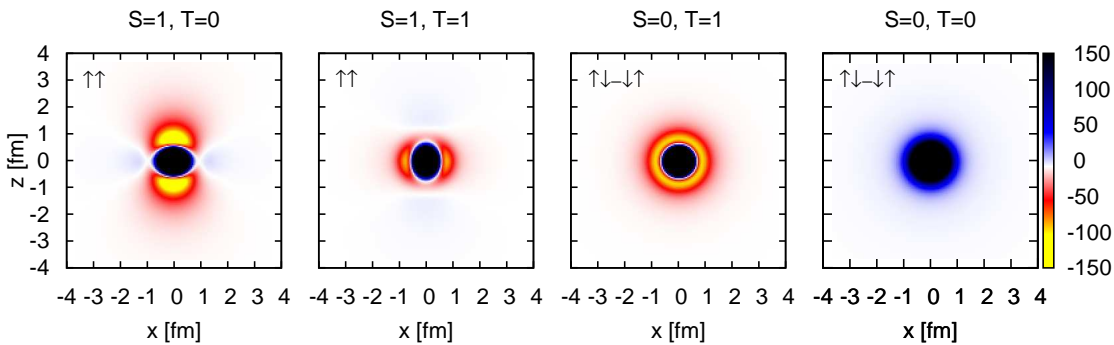}
  \caption{(Color online). Argonne v8$^{\prime}$ potential in the different
    spin-isospin channels as a function of the distance vector
    $\vec{r}=(x,y=0,z)$. In the $S=1$ channels the total spin is
    aligned with the $z$ axis. Units are in MeV.}
\label{fig:AV8pot}
\end{figure*}

The Argonne v8$^{\prime}$ (AV8$^{\prime}$) potential~\cite{wiringa95}
is depicted in Fig.~\ref{fig:AV8pot} as a function of $\vec{r}$ for the
four spin-isospin combinations of a nucleon pair.  In the left most
graph for $S=1, T=0$ we assume the nucleons to be at rest so that
the spin-orbit interaction does not contribute.  The tensor
interaction causes a quadrupole type dependence as a function of the
angle between the total spin direction (which we aligned along the
$z$ axis) and the direction of the distance vector $\vec{r}$.  The main
attraction is obtained when the spins of the nucleons are aligned with
the distance vector $\vec{r}$ while almost no attraction exists in the
$x$ direction where the spins are orthogonal to $\vec{r}$.  For $S=1,
T=1$ we added the spin-orbit interaction for $L_z=1$ because due to
the Pauli principle the nucleon pair has to be in an orbital state
with odd parity.  For $T=1$ the tensor interaction leads to a situation
which is opposite to the $T=0$ case.  Here the attraction occurs along
the $x$ axis where the spins are orthogonal to $\vec{r}$.

In the $S=0$ channels there is no tensor interaction and no spin-orbit
interaction thus the interaction depends only on the distance $|\vec{r}|$. Common to
all channels is the strong central repulsion for $|\vec{r}|<0.6$~fm. For $S=0, T=1$ there is strong attraction, around $|\vec{r}|=1$~fm, however, not strong enough to make the di-neutron bound. The $S=0, T=0$ potential is repulsive at all distances.

This rather complex nature of the nuclear interaction induces
corresponding intricate correlations in the $A$-body eigenstate of the
Hamiltonian which we discuss in Sec.~\ref{sec:results}.

\section{Results}
\label{sec:results}

In the following we investigate the ground states of \nuc{2}{H} with
$J^\pi=1^+$, \nuc{3}{H} and \nuc{3}{He} with $J^\pi=\frac{1}{2}^+$, and \nuc{4}{He}
with $J^\pi=0^+$, labeled by d, t, h, and $\alpha$, respectively, as
well as the excited $0^+$ state of \nuc{4}{He} at 20.21~MeV, which is a
resonance close to the threshold for \nuc{3}{H}~+~$p$, labeled by
$\alpha^*$. In this paper we treat the state $\alpha^*$ as a
quasi-bound state with a long tail~\cite{horiuchi08}, though it has a
proton width of 0.5~MeV.
 
\subsection{One-body densities}

\begin{figure}
  \includegraphics[width=0.85\columnwidth]{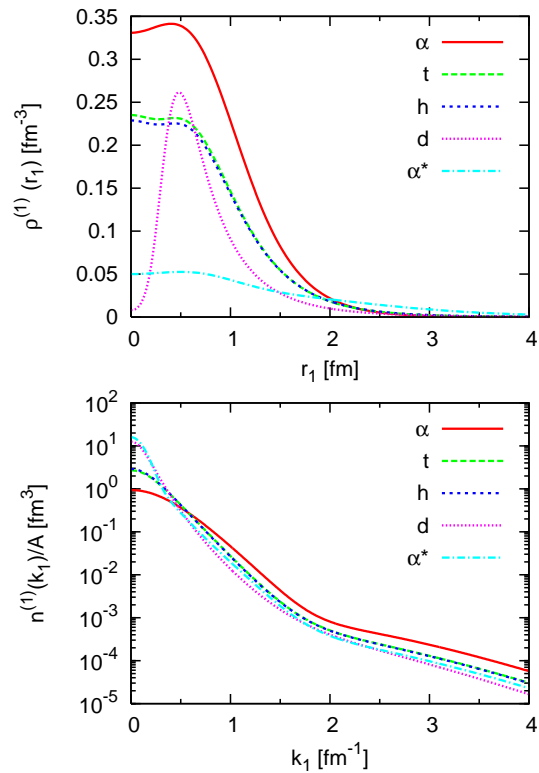}
  \caption{(Color online). One-body point densities of the different states in
    coordinate space (top) and one-body density in momentum
    space divided by mass number $A$ (bottom).  Ground states of
    \nuc{2}{H}, \nuc{3}{H}, \nuc{3}{He}, \nuc{4}{He} are denoted by d, t, h, and $\alpha$, respectively. The excited state of \nuc{4}{He} is labeled with
    $\alpha^*$.}
  \label{fig:dens1r}
\end{figure}

The one-body point densities of the five states are depicted in
Fig.~\ref{fig:dens1r}. In all cases the position $\vec{r}_1$ of the
nucleon is counted from the total center-of-mass position of the
many-body system. For the deuteron this means that $\vec{r}_1 = \frac{1}{2}\vec{r}$
is half the relative distance between neutron and proton.  The
densities are averaged over the directions of the total spin and hence
depend only on $r_1=|\vec{r}_1|$.  Likewise the momentum $\vec{k}_1$ of
a nucleon is the one in the total center of momentum frame and
averaging over total spin directions is implied.

The $\alpha$ particle shows the largest central density, the \nuc{3}{H} and
\nuc{3}{He} densities are somewhat smaller and differ only slightly due to
the Coulomb interaction.  The density of the excited $0^+$ state in \nuc{4}{He}
is much lower because this state, which is a narrow resonance in the
scattering of \nuc{3}{H} and proton, is essentially a configuration in
which a proton and a triton orbit around each other in an $l=0$
state~\cite{hiyama04,horiuchi08}.  Due to the recoil the quantal zero
point motion in the relative coordinate smears out the intrinsic
density of the triton.

We include also the deuteron despite the fact that its one-body
density is actually the two-body density at half the distance, and
only the $S=1, T=0$ component of the four possibilities to couple
spins and isospins of two nucleons contributes. The comparison with
the three- and four-body systems nicely demonstrates that in coordinate space
the effects of the short-range repulsion, which are clearly visible in
the deuteron, can not be seen in the $A$-body case because the
one-body density integrates over the positions of the other $A-1$
particles.

However the one-body momentum distribution (lower part of
Fig.~\ref{fig:dens1r}) shows beyond $k_1$~$\approx$~1.5~fm$^{-1}$ the presence
of short-range correlations by a far-out-reaching tail.  The
occurrence of high momenta is at variance with a Hartree-Fock like
mean-field picture where beyond the Fermi momentum $k_F\approx
1.4$~fm$^{-1}$ the momentum distribution drops
steeply~\cite{huefner81,alvioli05}.  For the deuteron the two-body density is
again identical to the one-body density (in momentum space
$\vec{k}_1 = - \vec{k}_2 = \vec{k}$).  One notices that the high momentum
tails have a very similar form in all cases including the deuteron.
This similarity suggests already a universal behavior of the
short-range correlations independent of the spatial density of
the $A$-body system.

\subsection{Two-body densities} 

\begin{figure*}
  \includegraphics[width=1.00\textwidth]{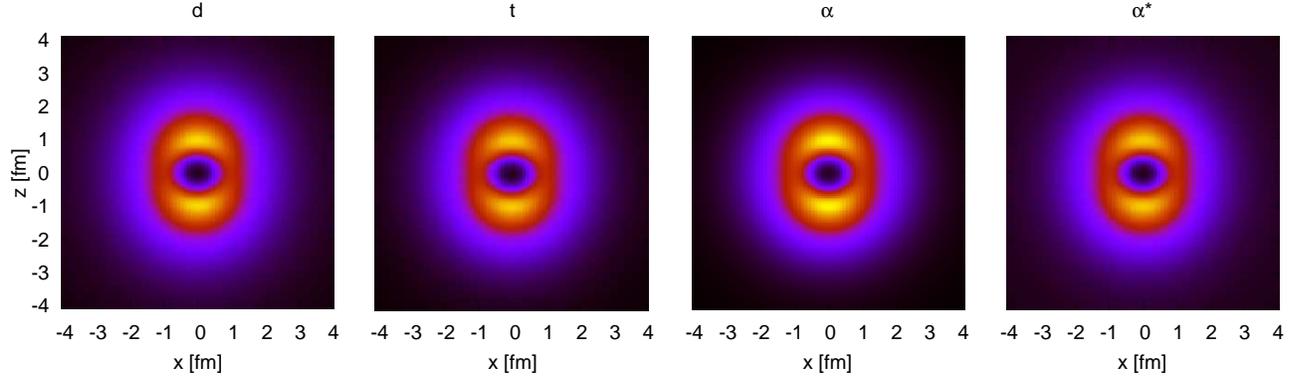}
  \caption{(Color online). From left to right: Two-body densities in coordinate space
    for a pair of nucleons with $S=1, M_S=1$ and $T=0$ in the
    ground states of \nuc{2}{H}, \nuc{3}{H} and \nuc{4}{He} and the
    20.21~MeV excited state of \nuc{4}{He} denoted by d, t, $\alpha$,
    and $\alpha^*$, respectively. The densities have rotational
    symmetry around the $z$ axis and range from black = 0 to
    bright (yellow) = maximum.  Maxima assume values of 0.008~fm$^{-3}$
    for d, 0.015~fm$^{-3}$ for t, 0.035~fm$^{-3}$ for $\alpha$, and
    0.015~fm$^{-3}$ for $\alpha^*$.
    }
  \label{fig:10r}
\end{figure*}

\begin{figure*}
  \includegraphics[width=0.765\textwidth]{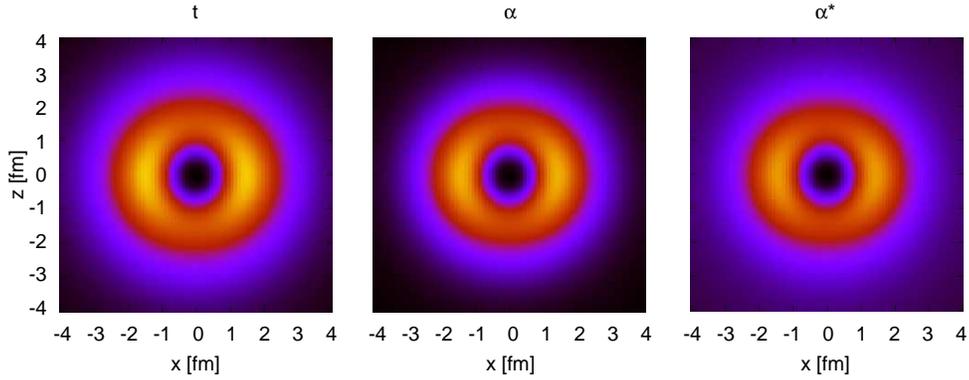}
  \caption{(Color online). From left to right: Two-body densities in coordinate space
    for a pair of nucleons with $S=1, M_S=1$ and $T=1$ of \nuc{3}{H}
    and \nuc{4}{He} in the ground states and the 20.21~MeV excited
    state of \nuc{4}{He} denoted by t, $\alpha$, and $\alpha^*$ respectively. The densities are axially symmetric around
    the $z$ axis and $M_T$ is summed over.  Otherwise same as
    Fig.~\ref{fig:10r}.  Maximum densities are
    $0.47\times10^{-3}$~fm$^{-3}$ for t, $2.2\times10^{-3}$~fm$^{-3}$
    for $\alpha$, and $0.51\times10^{-3}$~fm$^{-3}$ for $\alpha^*$.}
\label{fig:11r}
\end{figure*}

\begin{figure*}
  \includegraphics[width=0.765\textwidth]{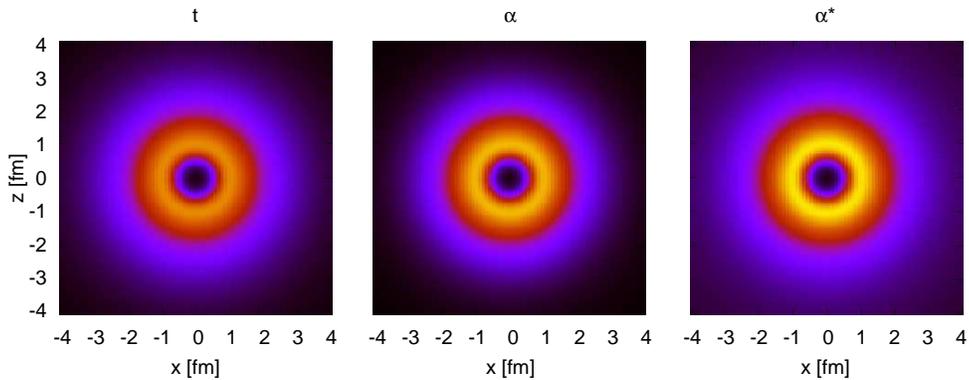}
  \caption{(Color online). The same as Fig.~\ref{fig:11r} but for a pair of nucleons
    with $S=0$ and $T=1$. Maximum densities are
    0.020~fm$^{-3}$ for t, 0.054~fm$^{-3}$ for $\alpha$, and
    0.019~fm$^{-3}$ for $\alpha^*$.}
  \label{fig:01r}
\end{figure*}

The $A$-body density, which contains the information about all
correlations, is a function of $A$ position or momentum vectors and
$4A$ spin-isospin possibilities and hence can not be visualized
easily. Therefore we integrate and sum over $A-2$ single-particle
degrees of freedom and are left with the two-body density.  This
represents the sum over all particle pairs in the many-body state.
In addition we integrate over the center-of-mass position of the pair
and obtain the two-body densities defined in Eqs.~(\ref{eq:rhorelr})
and (\ref{eq:rhorelk}) for the four spin-isospin channels which are
possible for a nucleon pair.  The complex nature of the
nucleon-nucleon interaction discussed in Sec.~\ref{sec:forces} induces
short-range repulsive and tensor correlations in the many-body state,
which can be seen best in the two-body density.

In Fig.~\ref{fig:10r} the spatial two-body densities
$\rho^{\rm{rel}}_{11,00}(\vec{r})$ of the four different states are
displayed. The first striking observation is that at short distances they look
very similar independently of the many-body state. That means that the
correlations felt by a particle pair in the $S=1, T=0$ channel are at
short distances the same independently of the remaining particles in
the system.  The second not unexpected observation is that these
densities reflect in an almost one-to-one fashion the potential in the
$S=1, T=0$ channel, see Fig.~\ref{fig:AV8pot}.
There exists a one-to-one correspondence between the nuclear Hamiltonian and the two-body densities. The expectation value of the Hamiltonian can be calculated with the two-body density as discussed in Ref.~\cite{suzuki09}.
In regions where the potential is most 
attractive, $\vec{r}\approx (0,0,\pm1$~fm$)$, the densities are large
and where the interaction is repulsive or close to zero the
probability of finding the particle pair is small.  At small distances
below 0.5~fm the AV8$^\prime$ potential is so strongly repulsive
that the pair densities in all many-body states are pushed down
toward zero.  One should bear in mind that in a simple shell model
many-body state these correlations can not be represented. The shell model
two-body densities have actually their maximum at relative distance
$\vec{r}=0$.

For $S=1, M_S=1, T=1$ the tensor interaction is most attractive in a torus
around the $z$ axis (see Fig.~\ref{fig:AV8pot}) and hence the two-body
density has its maximum in the $x$-$y$ plane as can be seen in
Fig.~\ref{fig:11r}. In this channel we see again a one-to-one
correspondence to the potential. For small distances up to about 1.5~fm the shape of the distribution is again almost identical for all three many-body states.
It should be noted that this channel, which does not exist for the deuteron, is weakly populated in the three- and four-body systems. Depending on the nucleus about 5--7\% of the pairs are in this channel, see Table~\ref{tab:pairs}.
In the shell model representation this channel corresponds to at least one
particle-hole excitation to the $p$ shell such that the relative
motion of the pair has negative parity. When occupying only the $s$
shell this channel is forbidden. We will discuss in Sec.~\ref{sec:threebody} how the two-body density in this channel is related to three-body correlations.

The second strong channel has $S=0, T=1$ and is shown in Fig.~\ref{fig:01r}.
As there is no tensor and spin-orbit interaction for $S=0$ the distributions are spherical.  Again they are very similar for all states and also exhibit a hole at short distances where the AV8$^\prime$ potential is very repulsive and a maximum at distances around 1~fm where it is most attractive, see Fig.~\ref{fig:AV8pot}.

\begin{table}
  \caption{Number of pairs in different states of light nuclei
    calculated with the AV8$^\prime$ potential. Calculated binding energy $E_b$ in MeV, matter point radius $\sqrt{\langle r^2 \rangle}$ in fm.}
\begin{ruledtabular}
\begin{tabular}{ccccccc}
state\textbackslash($ST$) & (10) & (01) & (11) & (00) & $E_b$ & $\sqrt{\langle r^2 \rangle}$\\
\hline
d           & 1   &  -- & --  & --  &$-$2.24&1.96\\
t           &1.490&1.361&0.139&0.010&$-$7.76&1.75\\
h           &1.489&1.361&0.139&0.011&$-$7.10&1.79\\
$\alpha$    &2.992&2.572&0.428&0.008&$-$25.09&1.49\\
$\alpha^*$  &2.966&2.714&0.286&0.034&$-$7.16&3.94\\
\end{tabular}
\end{ruledtabular}
\label{tab:pairs}
\end{table}

The $S=0, T=0$ channel is not displayed because its contribution listed in
Table~\ref{tab:pairs} is tiny, only about 0.1\%. The potential in this channel is purely repulsive as can be seen in Fig.~\ref{fig:AV8pot}. Nevertheless this small contribution is surprising when compared with the $S=1, T=1$ channel where the potential, while not purely repulsive, provides only very weak attraction. We will discuss this point later in relation to three-body correlations.

It is also interesting to note that the number of pairs in the $0_2^+$ state of the $^4$He nucleus are almost identical to the summed number of pairs from the triton and $^3$He --- reflecting the cluster nature of this state.  

\begin{figure}
  \includegraphics[width=0.85\columnwidth]{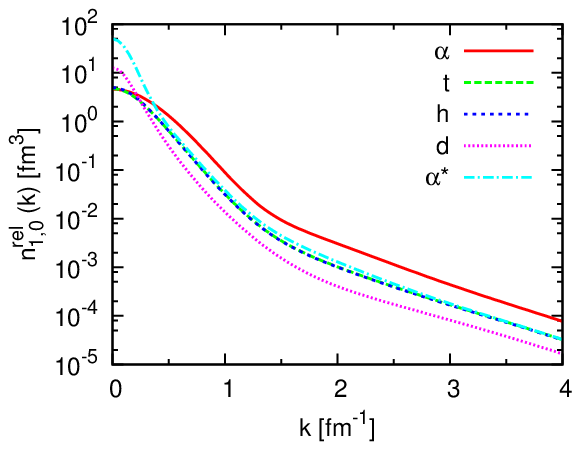}
  \caption{(Color online). Two-body densities as a function of relative momentum $k$
    for the $S=1, T=0$ channel. Ground state densities of \nuc{2}{H}, \nuc{3}{H},
    \nuc{3}{He}, \nuc{4}{He} are denoted by d, t, h, $\alpha$, respectively. The
    excited state of \nuc{4}{He} is labeled with $\alpha^*$.}
  \label{fig:dens2-10k}
\end{figure}

\begin{figure}
  \includegraphics[width=0.85\columnwidth]{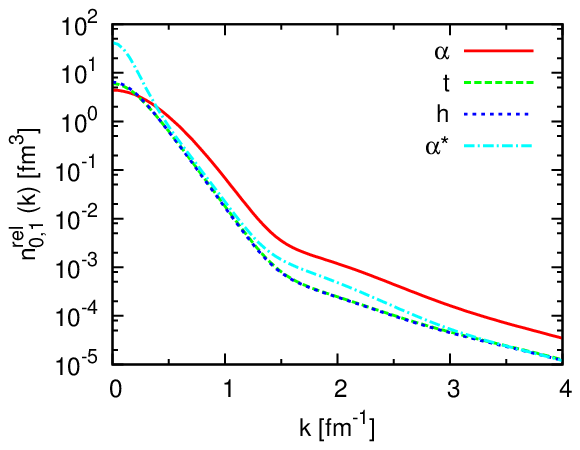}
  \caption{(Color online). The same as Fig.~\ref{fig:dens2-10k} but for the
    $S=0, T=1$ channel as a function of $k$.}
  \label{fig:dens2-01k}
\end{figure}

In Figs.~\ref{fig:dens2-10k} and \ref{fig:dens2-01k} we show the two-body densities in momentum space in the $S=1, T=0$ and $S=0, T=1$ channels. As expected we find more pairs at small relative momenta for the spatially extended deuteron and the excited $0^+$ state in \nuc{4}{He}. In all states we observe extended high momentum components above the Fermi momentum of about 1.4~fm$^{-1}$. Comparing both even channels we find that the two-body densities are very similar at low momenta up to about 0.5~fm$^{-1}$, but the high-momentum components in the $S=1, T=0$ channel are larger by a factor of 2-2.5. The differences in the number of pairs (see Table~\ref{tab:pairs}) in the even channels originates mainly from contributions at higher momenta between 0.5 and about 2.5~fm$^{-1}$. The larger number of high momentum pairs in the $S=1, T=0$ channel can be traced back to tensor correlations as we will discuss in Sec.~\ref{sec:ucom}.

These differences in the high-momentum contributions between the
$S=1, T=0$ and $S=0, T=1$ channels can also be interpreted in terms of proton-proton or neutron-neutron (only $T=1$) versus proton-neutron pairs (both $T=0$ and $T=1$). Such effects have been found also in theoretical studies for heavier nuclei \cite{schiavilla07,wiringa08,horiuchi07,alvioli08} and in experiment comparing for example the $(e,e'pp)$ with $(e,e'pn)$ cross sections where a dominance of proton-neutron pairs was observed \cite{subedi08}.

\subsection{Universality at small distances}

\begin{figure}
  \includegraphics[width=0.85\columnwidth]{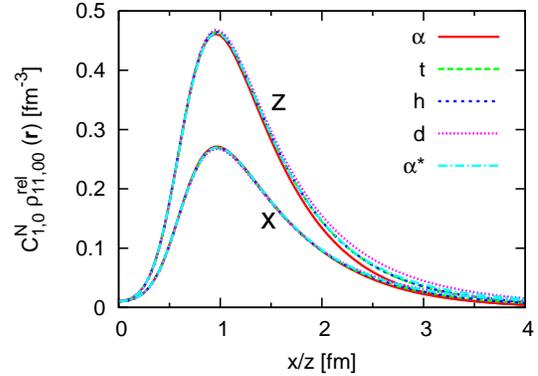}
  \caption{(Color online). Cuts of the normalized densities $\rho^{\rm{rel}}_{11,00}(\vec{r})$ for $\vec{r}=(x,0,0)$ and $\vec{r}=(0,0,z)$.}
  \label{fig:universality10r}
\end{figure}

As already seen in Figs.~\ref{fig:10r}-\ref{fig:01r} the two-body densities of the different states look very similar especially at small distances. To further investigate this universality of the short-range correlations we display in Fig.~\ref{fig:universality10r} cuts of the two-body density
$\rho^{\rm{rel}}_{11,00}(\vec{r})$ along the $z$- and the $x$-direction. As the absolute values of the densities are quite different in the five
states (see Fig.~\ref{fig:10r}) we normalize the two-body densities at $r=1$~fm, where the densities approximately reach their maximum value. The normalization factors
\begin{equation}
  C^N_{S,T} = \frac{1}{\rho^{\rm{rel}}_{S,T}(r = 1\,\mathrm{fm}) \, \mathrm{fm}^3}
\end{equation}
are given in Table~\ref{tab:normalization}. This choice for the normalization radius is not crucial as the ratios of the normalization coefficients between different states for a given channel are essentially constant (within 2\%) when calculated between 0 and 1.0~fm.
\begin{table}
	\caption{Normalization factors $C^N_{S,T}$.}
	\begin{ruledtabular}
	\begin{tabular}{ccccccc}
	 & d & t & h & $\alpha$ & $\alpha^*$ \\
	\hline
	$S=0, T=1$ & --    & 49.02 & 50.76 & 18.55 & 51.55 \\
	$S=1, T=0$ & 61.35 & 31.25 & 31.75 & 13.23 & 31.06 
\end{tabular}
\end{ruledtabular}
\label{tab:normalization}
\end{table}

\begin{figure}
  \includegraphics[width=0.85\columnwidth]{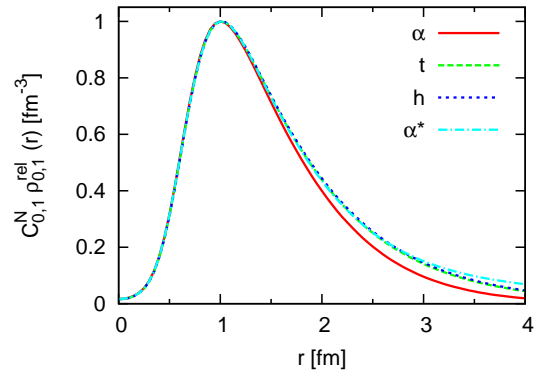}
  \caption{(Color online). Two-body densities $\rho^{\rm{rel}}_{0,1}(r)$ normalized to 1~fm$^{-3}$ at $r$=1~fm for different states (c.f. Fig.~\ref{fig:01r}).}
  \label{fig:universality01r}
\end{figure}

It is astonishing to see in Fig.~\ref{fig:universality10r} that for small
distances the scaled densities practically coincide along both the $z$ and the $x$ axes. This means that not only the central correlations but also the angular-dependence of the tensor correlations are almost identical at short distances.
The short-range central and tensor correlations exhibit universal
behavior at short distances below about 1~fm.

In the $S=0, T=1$ channel the same universal behavior can be observed as shown in Fig.~\ref{fig:universality01r}. The two-body densities normalized at $r=1\,\fm$ for the different systems agree perfectly to distances up to about 1~fm.

Whereas the behavior of the two-body densities at short distances is universal the behavior at large distances is specific for the particular many-body state. Its form is discussed in Ref.~\cite{suzuki09}.

\subsection{Universality at high momenta}

\begin{figure}
	\includegraphics[width=0.85\columnwidth]{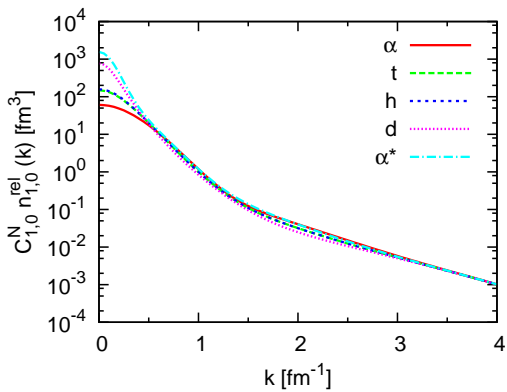}
  \caption{(Color online). Normalized two-body densities as a function of relative  momentum $k$ for the $S=1, T=0$ channel. Ground state densities of \nuc{2}{H}, \nuc{3}{H}, \nuc{3}{He}, \nuc{4}{He} are denoted by d, t, h, $\alpha$, respectively. The excited state of \nuc{4}{He} is labeled with $\alpha^*$.}
  \label{fig:dens2-10k-norm}
\end{figure}

\begin{figure}
  \includegraphics[width=0.85\columnwidth]{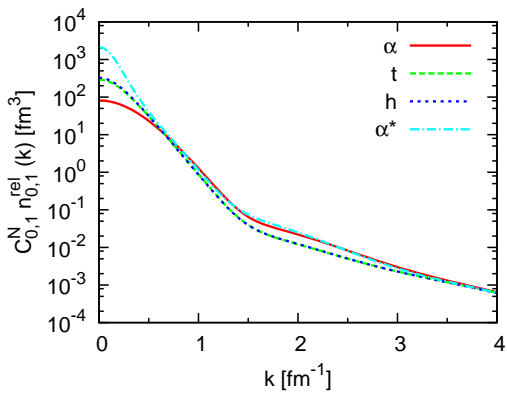}
  \caption{(Color online). The same as Fig.~\ref{fig:dens2-10k-norm} but for the
    $S=0, T=1$ channel as a function of $k$.}
  \label{fig:dens2-01k-norm}
\end{figure}

In Fig.~\ref{fig:dens2-10k-norm} and Fig.\ref{fig:dens2-01k-norm} we show the two-body densities in momentum space in the $S=1, T=0$ and $S=0, T=1$ channels scaled with the same normalization factors as given in Table~\ref{tab:normalization} that were determined for the two-body densities in coordinate space. Whereas the scaled densities differ strongly at low momenta, we find almost perfect agreement at high momenta larger than about 3~fm$^{-1}$. The universality of the short-range correlations in coordinate space is therefore reflected in a universality of the high-momentum components in momentum space. The fact that the two-body densities differ in the intermediate momentum range from the Fermi momentum of about 1.4~fm$^{-1}$ to about 3~fm$^{-1}$ should be related to differences in the long-range correlations for the different nuclei.

\subsection{Three-body correlations}
\label{sec:threebody}

\begin{figure}
	\includegraphics[width=0.7\columnwidth]{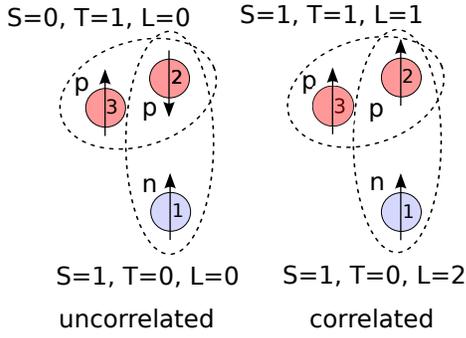}
	\caption{(Color online). Illustration of three-body correlations induced by tensor correlations. In the uncorrelated wave function (left) the two nucleons 1 and 2 are in an $S=1, M_S=0$ pair with $L=0$. The tensor force leads to an admixture of an $L=2$ component and an alignment of the spins of nucleons 1 and 2 flipping the spin of nucleon 2 (right). This affects the interaction between nucleon 2 and nucleon 3. In the uncorrelated wave function the protons 2 and 3 form an $S=0, T=1, L=0$ pair. After the spin-flip of nucleon 2 this becomes an $S=1, T=1, L=1$ pair.}
	\label{fig:threebodycorrelations}
\end{figure}

Looking at the number of pairs in the different spin-isospin channels (see Table~\ref{tab:pairs}) an interesting observation can be made. Let us concentrate on \nuc{4}{He}. In a simple shell model picture where all nucleons occupy $s$ orbits we should find three pairs each in the $S=1, T=0$ and $S=0, T=1$ channels and zero pairs in the odd channels. The nuclear potential is much more attractive in the even channels than in the odd channels, furthermore the kinetic energy is much higher in the odd channels due to the non-vanishing angular momentum. It is therefore surprising that we find in the exact wave function a remarkable depopulation of the $S=0, T=1$ even channel (2.572 pairs) obviously in favor of the $S=1, T=1$ odd channel (0.428 pairs). As remarkable is the fact that the number of pairs in the $S=1, T=0$ channel is essentially unchanged (2.992 pairs) compared to the simple shell model picture. This effect can not be understood in terms of two-body correlations, as the parity of the relative motion of a nucleon pair can not be changed by the two-body interaction. As already discussed by Forest \textit{et al.} \cite{forest96} this effect should be attributed to three-body correlations induced by the strong tensor force in the $S=1, T=0$ channel. As total isospin $T$ is a conserved quantity in light nuclei the total number of pairs in the $T=0$ and $T=1$ channels has to be conserved. The tensor force in the $S=1, T=0$ channel provides the dominant contribution to the nuclear binding. It has its origin in the pion exchange and is long ranged. Nucleon pairs in the $S=1, T=0$ channel will therefore be correlated even at large distances and these tensor correlations will affect other nucleon pairs. It is energetically favorable to break a pair in the $S=0, T=1$ channel by flipping the spin of a nucleon if this allows the tensor force to gain energy in another pair involving a third nucleon. An illustration of this mechanism is shown in Fig.~\ref{fig:threebodycorrelations} where energy is gained by tensor correlations for a pair of nucleons in the $S=1, T=0$ channel. In the uncorrelated case the nucleon pair is assumed to be in a relative $S$-wave. In the correlated many-body state the pair will be partially found in a relative $D$-wave to allow for additional binding by the tensor force. This $D$-wave admixture will also change the spin orientation of the nucleons, so that another pair, originally in the $S=0, T=1$ channel, is now found in the $S=1, T=1$ channel.  

\begin{figure}
	\includegraphics[width=0.85\columnwidth]{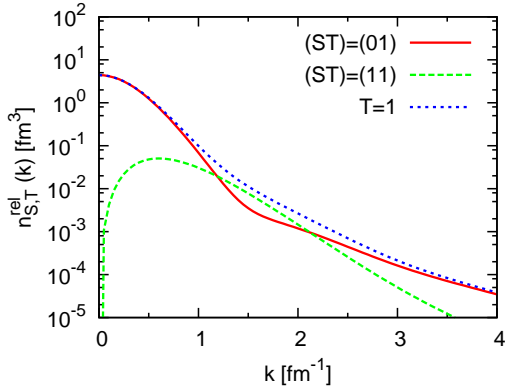}
	\caption{(Color online). Two-body densities in momentum space for \nuc{4}{He} in the $S=0, T=1$ and $S=1, T=1$ channels and the sum of both densities.}
	\label{fig:dens2k-T1}
\end{figure}

To illustrate the effects of these three-body correlation on the two-body densities in the $T=1$ channel we show in Fig.~\ref{fig:dens2k-T1} the two-body momentum distributions of the $S=0, T=1$ and the $S=1, T=1$ channels for \nuc{4}{He}.
At small relative momenta the density in the odd channel vanishes because of the $P$-wave nature. For momenta between 1.1 and 2.1~fm$^{-1}$ the two-body density in the $S=1, T=1$ is actually larger than in the $S=0, T=1$ channel. At very high relative momenta the contribution of the odd channel can again be neglected. The three-body correlations therefore influence the two-body density very differently in different momentum regimes. For low relative momenta below about 0.5~fm$^{-1}$ the effect is very small and the two-body densities in the two even channels are very similar. In an intermediate momentum range between 0.5 and 2.5~fm$^{-1}$ we observe a noticeable depletion of the $S=0, T=1$ channel in favor of the $S=1, T=1$ channel. This contributes to the fact that the two-body densities in the $S=1, T=0$ channel are much larger than in the $S=0, T=1$ channel in this momentum region.

As already mentioned this effect can not be understood in terms of two-body correlations. It also explains why effective interactions that are obtained by unitary transformations in two-body approximation, like $V_\mathit{low-k}$ \cite{bogner03}, the similarity renormalization group (SRG) \cite{bogner10} or the unitary correlation operator method (UCOM) \cite{ucom98,ucom03,ucom10}, provide more binding than the bare interaction when used in exact calculations. In two-body approximation the interaction is transformed independently in all spin-isospin channels. It is therefore possible to obtain the full contribution of the tensor force in the $S=1, T=0$ channel without having to pay the price of the three-body correlations. With increasing range of the tensor correlations (in the UCOM approach) or a lower cut-off (in the $V_\mathit{low-k}$ or SRG approaches) the effective interaction will induce smaller three-body correlations. Smaller three-body correlations means that less nucleon pairs are moved from the $S=0, T=1$ to the $S=1, T=1$ channel. As in the odd channel the potential is less attractive and the kinetic energy is much larger, the three-body correlations provide a repulsive contribution to the energy.

It has already been realized that a term in the effective interaction called antisymmetric spin-orbit (ALS)-force that connects $S=0$ with $S=1$ states and changes the relative angular momentum by $\Delta L$=1, like $(\vec{l}_1-\vec{l}_2)\cdot(\vec{\sigma}_1-\vec{\sigma}_2)$, is able to improve spectra and transition rates in $sd$-shell model calculations \cite{conze73,*feldmeier73}. But as such a term is not conserving translational and Galilei invariance it is not allowed in the free nucleon-nucleon interaction and can only be obtained by integrating many-body forces over additional particle degrees of freedom.

We want to stress the point that in our discussion no genuine three-body forces are considered. The three-body correlations are induced by the two-body tensor force. When genuine three-body forces are included we of course expect additional or modified three-body correlations.

\subsection{Comparison with unitary correlation operator method}
\label{sec:ucom}

The universality of short-range correlations is not only interesting in itself but also confirms the basic assumptions that underlie methods to derive effective low-momentum interactions such as UCOM, $V_{\mathit low-k}$ and SRG. We will discuss here the UCOM approach as it provides the most direct connection to the short-range correlations in the nucleus.

The basic idea of the UCOM approach is to imprint the short-range central and tensor correlations into the nuclear many-body wave functions explicitly by means of a unitary correlation operator $\op{C}$. Starting from an uncorrelated trial state $\ket{\Phi}$ the correlated state
\begin{equation}
  \ket{\Psi} = \op{C} \ket{\Phi}
\end{equation}
then features the short-range central and tensor correlations. Long-range correlations still have to be incorporated explicitly in the trial state \ket{\Phi}. 

To explain the action of the correlation operators we discuss first how the relative motion of two nucleons is affected by the correlation operators. For that we use basis states 
\begin{equation}
 \ket{\phi(LS)JM;TM_T} \: ,
\end{equation}
where the relative orbital angular momentum $L$ is coupled with the spin $S$ of the two nucleons to total angular momentum $J,M$. The isospin is coupled to $T,M_T$. The radial part of the relative wave function is given by $\phi(r)$.

In the $S=0$ channels only the central correlation operator acts and the correlated relative wave function is given, using the correlation function $\Rm(r)$, as
\begin{equation}
	\begin{split}
  \psi^{SJT}_{L}(r)
	& = \matrixe{r(LS)JT}{\CO_r}{\phi(LS)JT} \\
	& = \frac{\Rm(r)}{r}\sqrt{\DRm(r)}\; \phi(\Rm(r)) \ ,
  \end{split}
\end{equation}
whereas in the $S=1$ channels both central and tensor correlation operators act and we obtain the correlated radial wave functions
\begin{multline}
	\psi^{SJT}_{L;L'}(r) = \matrixe{r(L'S)JT}{\CO_{\Omega} \CO_r}{\phi(LS)JT} = \\
  \begin{cases}
    \frac{\Rm(r)}{r}\sqrt{\DRm(r)} \; \phi(\Rm(r)) & ; L'=L=J\\
    \cos \theta_J(r) \frac{\Rm(r)}{r}\sqrt{\DRm(r)} \; \phi(\Rm(r)) & ; L'=L=J\pm1\\
    \pm \sin \theta_J(r) \frac{\Rm(r)}{r}\sqrt{\DRm(r)} \; \phi(\Rm(r)) & ; L'=J\pm1,L=J\mp1
  \end{cases}
\end{multline}
with the tensor correlation function
\begin{equation}
 \theta_J(r) = 3 \sqrt{J(J+1)}\; \vartheta(r) \: .
\end{equation}

The functions $R_-(r)$ and $\vartheta(r)$ also carry implicitly the appropriate quantum numbers which are omitted here.

To calculate the two-body density in momentum space we will need the relative wave function in momentum space as obtained by Fourier transformation
\begin{multline}
 \matrixe{q(L'S)JT}{\CO_{\Omega}\CO_r}{\phi(LS)JT} = \\
	\sqrt{\frac{2}{\pi}} i^{L'} \int_0^\infty dr \, r^2
  j_{L'}(q r) \matrixe{r(L'S)JT}{\CO_{\Omega} \CO_r}{\phi(LS)JT} \: .
\end{multline}

To illustrate the action of the correlation operators we restrict the discussion here to the most simple trial state for \nuc{4}{He} where all nucleons occupy the $s$ orbit in a harmonic oscillator
\begin{equation}
 \label{eq:trialstate}
 \ket{\Phi} = \ket{(0s)^4} \: .
\end{equation}
The harmonic oscillator width parameter $a=1.98\,\fm^2$ is adjusted to reproduce the radius of the \nuc{4}{He} nucleus as obtained in the exact calculation with the correlated Gaussian approach.

We can then express the uncorrelated two-body density operator for this state as
\begin{equation}
	\begin{split}
	\op{\rho}^{(2)}_{\mathrm{uncorr}} =
	& \sum_{M_T} \ket{\phi_0(00)0;1M_T}\bra{\phi_0(00)0;1M_T} + \\
	& \sum_{M} \ket{\phi_0(01)1M;00}\bra{\phi_0(01)1M;00}
	\end{split}
\end{equation}
with the relative $L=0$ wave function
\begin{equation}
 \phi_0(r) = \left( \frac{2}{\pi a^3} \right)^{1/4} \exp\left\{- \frac{r^2}{4 a} \right\} \: .
\end{equation}

Including short-range central and tensor correlations with the UCOM correlation operators the two-body density operator of the correlated state is given in two-body approximation as
\begin{equation}
	\label{eq:corrtbd}
	\begin{split}
	\op{\rho}^{(2)} 
	= & \: \CO_{\Omega}\CO_r \: \op{\rho}^{(2)}_{\mathrm{uncorr}} \: \COd_r \COd_{\Omega} \\
	= & \sum_{M_T} \ket{\psi^{001}_{0}(00)0;1M_T}\bra{\psi^{001}_0(00)0;1M_T} + \\
	  & \sum_{M} \biggl( \ket{\psi^{110}_{0;0}(01)1M;00}+\ket{\psi^{110}_{0;2}(21)1M;00} \biggr) \times \\
		& \qquad \biggl( \bra{\psi^{110}_{0;0}(01)1M;00}+\bra{\psi^{110}_{0;2}(21)1M;00} \biggr) \ ,
	\end{split}
\end{equation}
where $L=2$ components appear in the $S=1, T=0$ channel due to the tensor correlation operator.

In the following we use correlation functions derived from an SRG evolved AV8$^\prime$ Hamiltonian \cite{hergert07,ucom10}. The used flow parameters $\alpha$ = 0.04~fm$^4$ and $\alpha$ = 0.20~fm$^4$ correspond to cut-off parameters of $\lambda$ $\approx$ 2.2~fm$^{-1}$ (soft) and $\lambda$ $\approx$ 1.5~fm$^{-1}$ (very soft). The labels UCOM04 and UCOM20 will be used in the following to identify these two sets of correlation functions. In exact calculations using the no-core shell model \cite{ucom10} the corresponding UCOM interactions provide binding energies that are close to the experimental binding energies for \nuc{3}{H} and \nuc{4}{He}. Using the simple trial state in Eq.~(\ref{eq:trialstate}) we obtain \nuc{4}{He} binding energies of $-$18.50~MeV and $-$25.10~MeV with UCOM04 and UCOM20, respectively.

\begin{figure}
  \includegraphics[width=0.85\columnwidth]{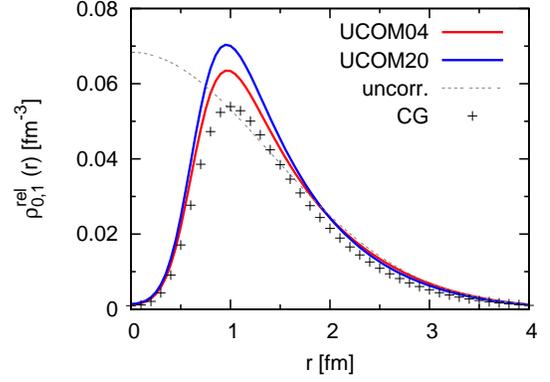}
	\caption{(Color online). Comparison of the coordinate space two-body density in the $S=0, T=1$ channel in \nuc{4}{He} between the UCOM and the exact many-body calculation using correlated Gaussians denoted by CG. See the text for the different UCOM and uncorrelated results.}
	\label{fig:ucomdensr01}
\end{figure}

\begin{figure}
  \includegraphics[width=0.85\columnwidth]{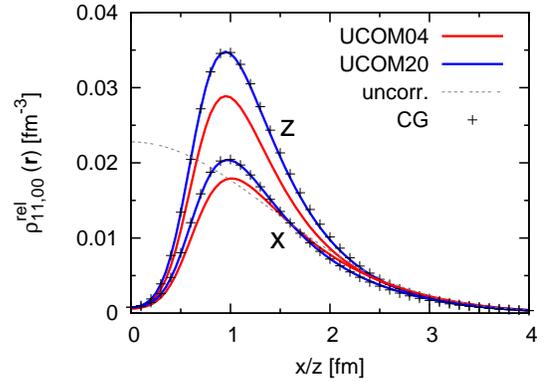}
	\caption{(Color online). The same as Fig.~\ref{fig:ucomdensr01} but for the two-body density in the $S=1, T=0$ channel as a function of $x$ and $z$.}
	\label{fig:ucomdensr10}
\end{figure}

In Figs.~\ref{fig:ucomdensr01} and \ref{fig:ucomdensr10} we compare for the two even channels the two-body densities in coordinate space given by Eq.~(\ref{eq:corrtbd}) with the two-body densities calculated from the exact solution for \nuc{4}{He}. The two-body densities obtained within the UCOM approach agree very well with the exact two-body densities at small distances. Compared with the uncorrelated wave function the two-body density is strongly suppressed at short distances. At $r$ around 1~fm the UCOM two-body densities depend on the specific choice of the correlation functions and deviate by 0--20\% from the exact results. In the $S=0, T=1$ channel the UCOM two-body densities are always larger than the exact results. The main reason for this discrepancy is the two-body approximation that we used to obtain the correlated two-body density. The UCOM results do not include the effects of three-body correlations and the number of pairs in the presented UCOM result is exactly three in both even channels.

In the $S=1, T=0$ channel (Fig.~\ref{fig:ucomdensr10}) we find an almost perfect agreement of the exact two-body densities with the UCOM20 result which uses long-range correlation functions. This holds not only for the radial dependence due to short-range repulsion but also for the angular dependence of the two-body density due to the tensor correlations. For the shorter-ranged correlation functions UCOM04 the agreement is not so good. This is caused by the different ranges of the tensor correlation functions. It appears that the long-range correlation functions in UCOM20 are able to describe most of the tensor correlations found in \nuc{4}{He} whereas with the short-ranged tensor correlation functions in UCOM04 a significant part of the medium to long-range tensor correlations is missing.

\begin{figure}
  \includegraphics[width=0.85\columnwidth]{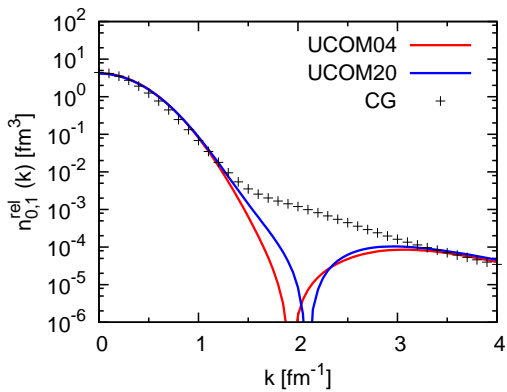}
	\caption{(Color online). UCOM two-body density in momentum space for the $S=0, T=1$ channel in \nuc{4}{He} compared with exact many-body calculation denoted by CG.}
	\label{fig:ucomdensk01}
\end{figure}

\begin{figure}
  \includegraphics[width=0.85\columnwidth]{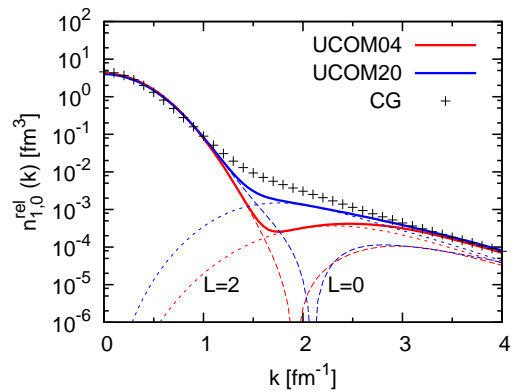}
	\caption{(Color online). The same as Fig.~\ref{fig:ucomdensk01} but for the $S=1, T=0$ channel. Contributions from the $L=0$ and $L=2$ components are shown for the UCOM densities.}
	\label{fig:ucomdensk10}
\end{figure}

The two-body densities in momentum space obtained with the UCOM densities given in Eq.~(\ref{eq:corrtbd}) are compared with the exact results in Fig.~\ref{fig:ucomdensk01} and Fig.~\ref{fig:ucomdensk10}. The effect of the short-range correlations are reflected in the high-momentum components. For relative momenta larger than about 3~fm$^{-1}$ we find good agreement with the exact result for both UCOM transformations. In the $S=0, T=1$ channel the UCOM densities are much too small in the intermediate momentum region from 1.4 to 3~fm$^{-1}$. This reflects the oversimplified Gaussian trial wave function and missing many-body correlations.

For the two-body densities in the $S=1, T=0$ channel (Fig.~\ref{fig:ucomdensk10}) we have decomposed the UCOM results in the $L=0$ and the $L=2$ components. The $L=0$ component looks very similar to the two-body densities obtained in the $S=0, T=1$ channel and does not contribute in the intermediate momentum-region. The $L=2$ component, introduced by the tensor correlation operator, on the other hand actually dominates the medium- and high-momentum part of the two-body density. There is a strong dependence on the range of the correlation functions but even in the UCOM20 case the exact two-body densities are still significantly larger in the intermediate momentum region. Again, contributions due to long-range correlations are missing.

To include these missing contributions consistently the UCOM two-body densities should be calculated not from the simple trial state in Eq.~(\ref{eq:trialstate}) but from an exact solution $\ket{\Phi}$ of the many-body problem
\begin{equation}
	\op{H}_\mathrm{UCOM} \ket{\Phi} = E \ket{\Phi}
\end{equation}
using the UCOM effective Hamiltonian in two-body approximation
\begin{equation}
 \op{H}_\mathrm{UCOM} = \op{C}^\dagger \op{H} \op{C} \: .
\end{equation}
Such calculations are in preparation using the no-core shell model. Nevertheless we can not expect perfect agreement even in this case due to the two-body approximation. Using an effective interaction like UCOM the three-body correlations as discussed in Sec.~\ref{sec:threebody} will not be fully included. The deviations between the exact two-body densities and the two-body densities obtained using effective interactions in two-body approximations will depend on the range of the correlation operators (in the UCOM approach) or on the value of the cut-off (in SRG and $V_\mathit{low-k}$).

A detailed discussion of the operator evolution in the SRG for the deuteron is provided by Anderson \textit{et al.} \cite{anderson10}. The authors study the evolution of high-momentum operators like the momentum distribution within the SRG and investigate to what extend a decoupling between low- and high-momentum components occurs. 

\section{Conclusions}
\label{sec:summary}


In this paper we have studied the two-body densities in coordinate and momentum space for the deuteron, \nuc{3}{H}, \nuc{3}{He}, \nuc{4}{He} and the first excited $0^+$ state in \nuc{4}{He}. Fully converged solutions for these light nuclei could be achieved using the correlated Gaussian basis approach for the Argonne~v8$^\prime$ interaction. The short-range repulsion and the tensor force induce strong short-range correlations in the many-body wave functions, reflected in the two-body densities. If the two-body densities in coordinate space are normalized at short distances, we find in the different spin-isospin channels a universal behavior up to about 1~fm in all nuclei. Using the same normalization we observe a corresponding universal behavior of the two-body densities in momentum space at relative momenta larger than about 3~fm$^{-1}$. Although we only have two-body forces we could identify three-body correlations due to the long-range tensor correlations in the $S=1, T=0$ channel. They manifest themselves in the two-body densities by a reshuffling of pairs from the  $S=0, T=1$ channel into the $S=1, T=1$ channel. 

The universal behavior of the short-range correlations explains the success of approaches such as  $V_\mathit{low-k}$ \cite{bogner03}, or SRG \cite{bogner10} and UCOM \cite{ucom10} that use unitary transformations to derive an effective low-momentum interaction. The idea of the unitary transformation is to decouple the short-range from the long-range or the high-momentum from the low-momentum physics. Using such transformed low-momentum interactions the wave functions no longer show the strong short-range correlations induced by the original interaction. To recover short-range correlations the two-body densities have to be transformed using the same unitary transformation. We compared in coordinate and momentum space the exact two-body densities of \nuc{4}{He} to those obtained from a simple $0\hbar\Omega$ trial wave function and the UCOM transformation for the Argonne~v8$^\prime$ interaction. In the $S=1, T=0$ channel we find a very good agreement for the short-range and the high-momentum behavior of the two-body densities. Differences show up mostly in the intermediate momentum range from 1.5 to 3~fm$^{-1}$. In this region long-range correlations, missing in the very simple trial wave function, become important. In the $S=0, T=1$ channel the agreement is spoiled by the missing three-body correlations in the UCOM approach. To recover these differences the unitary transformation would have to be performed not in a two-body approximation as done here but on the three-body level. 
  

In a more elaborate approach short-range correlations in heavier nuclei will be studied by solving the many-body problem with a soft unitarily transformed interaction for example with the no-core shell model \cite{ucom10} and then calculating the unitarily transformed two-body densities. This will allow to include both, long-range correlations by the many-body approach and short-range correlations by the unitary transformation. In the two-body approximation the role of three-body correlations could be investigated by varying the cut-off of the transformation. An explicit treatment of three-body correlations is possible in principle but would become very involved. 

We studied the two-body densities in this paper only as a function of the distance or the relative momentum of the nucleons, but it would also be interesting to investigate the dependence on the center of mass momentum of the nucleon pairs. Wiringa \textit{et al.} found that the short-range correlation effects are most pronounced at vanishing center-of-mass momentum for the pairs \cite{wiringa08}. For larger center-of-mass momentum the short-range correlations are smeared out as there is a higher probability to find one nucleon inside the Fermi sphere even at high relative momentum. It might also be interesting to study the two-body densities as a function of the center-of-mass position. In particular it might be possible to study short-range correlations of neutrons in the surface of neutron-rich exotic nuclei. Because of the universality of the short-range correlations information from this low-density regime should also be important for the saturation properties of neutron matter at higher densities. Of course three-body forces will become more and more important with increasing density.

\begin{acknowledgments} 
	We acknowledge support by the ExtreMe Matter Institute EMMI and HIC for FAIR. W.~H. is supported by the Special Postdoctoral Researchers Program of RIKEN. Y.~S. is supported in part by a Grant-in-Aid for Scientific Research (No.~21540261).
\end{acknowledgments}

\bibliography{src}

\end{document}